\documentclass[10pt,letterpaper]{article}
\usepackage{opex3}
\usepackage{verbatim}

\begin{document}

\title{Phase effects due to beam misalignment on diffraction gratings}

\author{Deepali Lodhia$^{1^{\ast}}$, Daniel Brown$^1$, Frank Br\"uckner$^1$, Ludovico Carbone$^1$, Paul Fulda$^2$, Keiko Kokeyama$^3$ and Andreas Freise$^1$}

\address{$^1$School of Physics and Astronomy, University of Birmingham, Edgbaston,\\ Birmingham, B15 2TT, UK}
\address{$^2$Department of Physics, University of Florida, Gainesville,\\ Florida, 32611-8440, USA}
\address{$^3$Department of Physics and Astronomy, Louisiana State University, Baton Rouge,\\ Louisiana, 70803-4001, USA}

\email{$^{\ast}$dl@star.sr.bham.ac.uk} 



\begin{abstract}
All-reflective interferometer configurations have been proposed for the next generation of gravitational wave detectors, with diffractive elements replacing transmissive optics. However, an additional phase noise creates more stringent conditions for alignment stability. A framework for alignment stability with the use of diffractive elements was required using a Gaussian model. We successfully create such a framework involving modal decomposition to replicate small displacements of the beam (or grating) and show that the modal model does not contain the phase changes seen in an otherwise geometric planewave approach. The modal decomposition description is justified by verifying experimentally that the phase of a diffracted Gaussian beam is independent of the beam shape, achieved by comparing the phase change between a zero-order and first-order mode beam. To interpret our findings we employ a rigorous time-domain simulation to demonstrate that the phase changes resulting from a modal decomposition are correct, provided that the coordinate system which measures the phase is moved simultaneously with the effective beam displacement. This indeed corresponds to the phase change observed in the geometric planewave model. The change in the coordinate system does not instinctively occur within the analytical framework, and therefore requires either a manual change in the coordinate system or an addition of the geometric planewave phase factor.
\end{abstract}



\section{Introduction}
A network of ground-based laser interferometric observatories,
including Advanced LIGO \cite{harry2010advanced}, Advanced VIRGO
\cite{accadia2011status} and GEO HF \cite{willke2006geo}, are in the
midst of being upgraded to highly sensitive second-generation
gravitational wave detectors. Scientists are confident of obtaining
the first direct detection of gravitational waves using the modified
instruments, based on the experience gained through the
first-generation operation. A simultaneous effort is also underway to
devise new concepts to increase detector sensitivities even further,
in an attempt to introduce a new era of gravitational-wave astronomy.

One promising approach for the next-generation of interferometers is
to replace conventional partly-transmissive optics, such as beam
splitters and cavity couplers, by reflective dielectric diffraction
gratings \cite{drever1996concepts}. An all-reflective interferometer
setup has the potential to significantly reduce thermal effects 
caused by absorption of high-power laser light in the optical
substrates. These thermal distortions are known as a significant challenge
limiting the light power that can be utilized in current and future interferometer
gravitational wave detectors. Additionally, reflective diffractive
gratings allow for a broader choice of opaque substrate materials with
a potential for superior mechanical properties.

\begin{figure}[htb]
\centering
\begin{minipage}[c]{.48\textwidth}
\centering
\includegraphics[width=8.3cm]{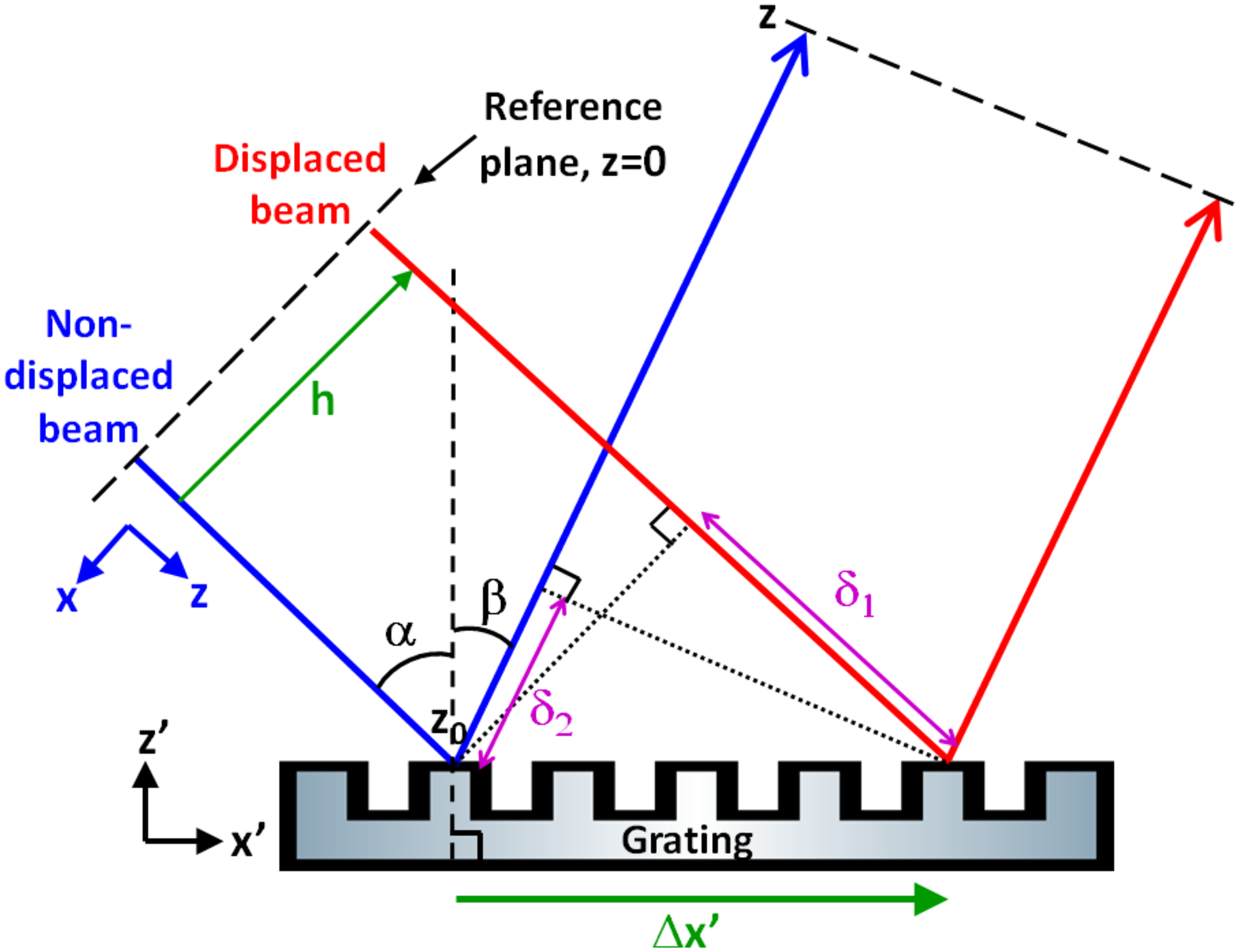}
\end{minipage}
\hspace{0.35cm}
\begin{minipage}[c]{.48\textwidth}
\centering
\includegraphics[width=4.6cm]{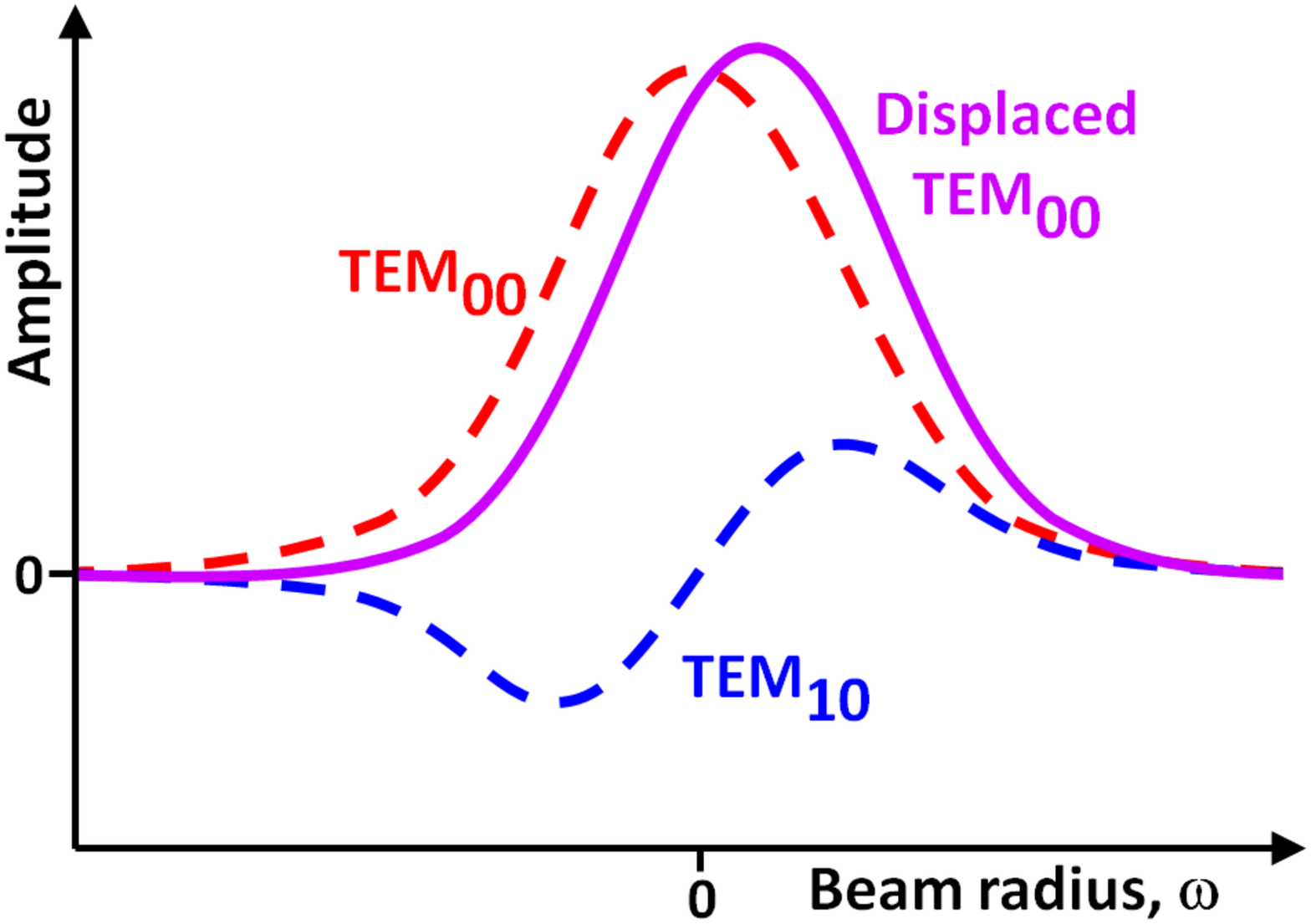}
\end{minipage}
\caption{\emph{Left}: Diffraction of light into the $m$-th diffraction
  order when the grating is displaced by amount $\Delta x'$ relative
  to the beam. A grating displacement $\Delta x'$ corresponds to a
  parallel beam displacement $h$, and induces an output optical path
  length change of $\Delta P$ according to
  Eq.~(\ref{eq:noise}). \emph{Right}: A displaced zero-order mode beam
  can be decomposed into non-displaced zero-order and first-order mode
  beams for fixed coordinate systems.}
\label{geometry}
\end{figure}

Since the initial concept of all-reflective interferometry, various
groups around the world have built on the notion and presented
experimental proof of their feasibility and compatibility with
standard interferometric techniques. Such work includes replacing
conventional cavity input mirrors and characterising 2-port
\cite{bunkowski2006diffractive} and 3-port cavity couplers
\cite{bunkowski2004low,britzger2011pound}, as well as a full operation
of a 4-port reflectively coupled Michelson interferometer
\cite{friedrich2008diffractive}. A very recent proposal also
demonstrated reflective coupling without the need of adding a
multilayer coating, which would further reduce thermal effects
associated with the coating stack itself
\cite{kroker2011reflective}. However, due to the broken symmetry of
light deflection, the use of diffraction gratings introduces an
additional coupling from alignment noise to output phase noise
\cite{wise2005phase,freise2007phase}.

Figure~\ref{geometry} (left) illustrates how the phase noise arises
from a simple misalignment of a beam. An incident beam with vacuum
wavelength $\lambda$ is diffracted from a reflective diffraction
grating with period $d$ into the $m$-th diffraction order (we consider
only one diffraction order $m$). The angle conventions imply a
positive incidence angle $\alpha$, while the diffraction angles
$\beta_m$ can be both negative and positive. According to the grating
equation

\begin{equation}
\textrm{sin} \, \alpha + \textrm{sin} \, \beta_{m} = \frac{m\lambda}{d},
\label{eq:grating}
\end{equation}

it follows that the diffraction angle $\beta_m$ of a certain
functional output coupling port will be different from the incident
angle $\alpha$ (except for the zero-order specular order). It is
discernible from a purely geometrical consideration that a slight
lateral displacement of either the grating, $\Delta x'$, or the beam,
$h$, induces a shift in the optical path length of $\Delta P =
\delta_1 + \delta_2$ (where $\delta_2$ is negative). This shift is
related to the displacement $\Delta x'$ via the grating equation
(\ref{eq:grating}) \cite{freise2007phase}:

\begin{equation}
\Delta P = \delta_1 + \delta_2 = - \Delta x' \frac{m\lambda}{d}.
\label{eq:noise}
\end{equation}

In other words, when a grating is displaced by an amount $\Delta x=d$,
a diffracted beam will undergo a total phase shift of $2\pi$ radians,
with a dependency on the diffraction order $m$. Given this phase
noise, the use of gratings imposes more challenging requirements for
the suspension and isolation systems for optical components compared
to conventional interferometric settings with a natural symmetry of
light reflection. Recently, we proposed an advanced readout for the
ports of a grating coupler which suppressed phase noise originating
from lateral grating displacement. The result was a factor of 20
relaxation in the lateral displacement suspension requirement at
10\,Hz \cite{hallam2009coupling}.

There is an urgent need to assess these requirements in further detail
with the use of realistic interferometry simulation tools based on a
Gaussian description, such as \textsc{Finesse}
\cite{freise2004frequency}. These Gaussian-based simulation tools are
restricted to fixed coordinate systems, and therefore rely on the
technique of decomposing the laser beam into higher-order modes. Yet,
effects due to grating-related phase noise were solely investigated
using geometric planewave considerations because they allow for a
straight-forward computation of the phase shift; these plane-wave
models are not a realistic representation of the Gaussian-based
simulation tools.

In this paper, we present a fully analytical Gaussian framework to
describe a small beam (or grating) displacement by means of a
first-order modal decomposition of the beam. This modal model is the
standard approach for off-axis beams in Gaussian-based simulation
tools, as depicted in Fig.~\ref{geometry} (right). We examine the
phase behaviour of the displaced beam and compare the results for the
modally decomposed model with those from the geometric planewave
approach (preliminary studies and results can be found in
\cite{lodhia2012phase}). We justify the use of modal decomposition
within the Gaussian framework through experimental means: we
investigate the phase behaviour in a zero-order and first-order beam
imprinted by diffraction from a grating in order to determine
whether the phase is dependent on the beam shape. The experimental
setup consisted of a table-top Mach-Zehnder interferometer, and a
reflective diffraction grating placed in one of the arms. Using a
Gaussian description as the core basis, a rigorous time-domain
simulation tool was developed to firstly verify the phase dependency
on the beam shape, thereby further supporting the experimental
findings. Secondly, a direct phase comparison of a diffracted beam was
made between an actual beam displacement and a modally decomposed
model. We finally corroborate the inclusion of the `geometric
planewave phase factor' for alignment stability computations involving
diffractive elements when using Gaussian-based simulation tools.

\section{A framework for a Gaussian beam displacement}
\label{sec:theory}
The aim of this section is to establish an analytical framework
derived from a Gaussian description of modal decomposition. Using the
phase distribution of a non-displaced beam as a reference, we then
distinguish the phase accumulation of a displaced beam using two
approaches: an actual geometrical translation of the beam or grating,
and modal decomposition by adding a higher-order mode.

To clarify the setup, we refer to Fig.~\ref{geometry} (left), where a
non-displaced and displaced beam both propagate from one reference
plane (black dashed line) and undergo grating diffraction before
reaching a second reference plane. The grating is orientated to lie in
the $x'$-$y'$ plane, with the grooves parallel to the $y'$-axis. The
beam propagates along the $z$-axis, and we are concerned with changes
to the beam parameters only in the $x$-$z$ direction of the beam. Note
that the coordinate system of the beam ($x,y,z$) is rotated by the
angle $\alpha$ relative to the coordinate system of the grating
($x',y',z'$). The grating displacement $\Delta x'$ is expressed in
terms of the beam displacement $h$ and the angle of incidence $\alpha$
using the following relation:
\begin{equation}
\label{xh}
h = \Delta x' \, \textrm{cos} \, \alpha.
\end{equation}

\subsection{Beam translation and modal decomposition of a Gaussian beam}
\label{subsec:modal}
We begin by considering the description of a displaced Gaussian beam
in terms of both a geometric translation and a modal decomposition
(see Appendix A for a thorough description of a Gaussian
beam). Without loss of generality, we initially consider the beam at
the waist position, $z_0$, where the waist size is $\omega_0$. We also
assume a displacement of the beam along the $x$-axis, rather than a
displacement of the grating along the $x'$-axis, due to the symmetry
of the setup~\cite{wise2005phase}. If we introduce a displacement $h$
to a fundamental beam, the Hermite-Gauss function of the translated
beam, $u_0^t(x,z_0)$, is defined at the waist as:

\begin {equation}
\label{eqn7}
u_0^t(x,z_0) = \left(\frac{2}{\pi}\right)^{\frac{1}{4}} \frac{1}{\sqrt{\omega_0}} \textrm{exp} \left(- \frac{(x-h)^2}{\omega_0^2}\right),
\end{equation}

where the superscript $t$ indicates a geometric translation along the
$x$-axis. Next, we can substitute Eq.~(\ref{eqn4}) into an expansion
of Eq.~(\ref{eqn7}) (see Appendix A). Since typical grating
displacements are extremely small compared to the beam waist, we can
use the approximation $h/\omega_0 \ll 1$ and apply a first-order
Taylor expansion to obtain the expression

\begin{equation}
\label{eqn8}
u_0^d(x,z_0) = u_0(x,z_0) + \frac{h}{\omega_0} u_1(x,z_0),
\end{equation}

with the superscript $d$ to denote a first-order modal
decomposition. Equation~(\ref{eqn8}) validates the theorem that the
properties of a slightly displaced zero-order mode beam can be
characterised by a decomposition into zero-order and first-order modes
(as illustrated in Fig.~\ref{geometry} (right)).

\subsection{Phase terms}\label{phaseterms}
Having established an expression to describe a displaced beam by means
of modal decomposition, the phase of the decomposed beam can be
examined more closely. The specific phase terms can be better
understood when the beam is propagated away from the waist, and $z_0$
is replaced by the propagation distance along the optical axis,
$z$. Away from the waist, extra parameters have to be taken into
consideration, such as the radius of curvature of the wavefronts,
$R_C$, the Gouy phase, $\Psi(z)$, and the wave number $k$ (defined as
$k=2\pi/\lambda$). Equations~(\ref{eqn1}) and (\ref{eqn3}) in Appendix
A reveal three contributions to the overall phase of a beam:
$\textrm{exp}(-ikz)$, $\textrm{exp}\left(i\frac{1}{2}\Psi\right)$ and
$\textrm{exp}\left(-i\frac{kx^2}{2R_C}\right)$. Using the general
form, $\textrm{exp}(-i\theta)$, the phase of a beam, $\theta$, at any
given point in the $x$-$z$ plane is specified as

\begin{equation}
\label{eqn9}
\theta_{f,t,d} = kz - \frac{1}{2}\Psi + \phi_{f,t,d},
\end{equation}

where the subscripts $f$, $t$ and $d$ correspond to the fundamental
non-translated, translated and modally decomposed beams,
respectively. The individual terms for $\phi_{f,t,d}$ are detailed in
Appendix B. We now have a Gaussian-based framework to exactly describe
the phase of non-displaced and displaced beams before any grating
diffraction. Next, we introduce diffraction due to a grating and
explore the impact on the phase terms.

\subsection{Effects from astigmatism upon diffraction}
The effect of diffraction into various angles which are different
from the angle of incidence can be accounted for
by introducing an \emph{astigmatism} to the Gaussian beam. As demonstrated in
Fig.~\ref{geometry} (left), a beam incident on the grating will
possess different angles of incidence and reflection, i.e. $\alpha
\neq \beta$. Consequently, the diffracted beam produces an elliptical
beam spot, only the beam parameters along the $x$-axis change, whilst
those along the $y$-axis remain the same. This astigmatism results in
a different waist size along the $x$-axis in the diffracted beam,
$\omega_{0_r}^x$, which can be expressed in terms of the waist size of
the incident beam, $\omega_{0_i}$, using the following relation

\begin{equation}
\label{eqn14}
\omega_{0_r}^x = \omega_{0_i} \frac{\textrm{cos}\,\beta}{\textrm{cos}\,\alpha}.
\end{equation}

\begin{figure}[htb]
\centering
\includegraphics[width=4.3in]{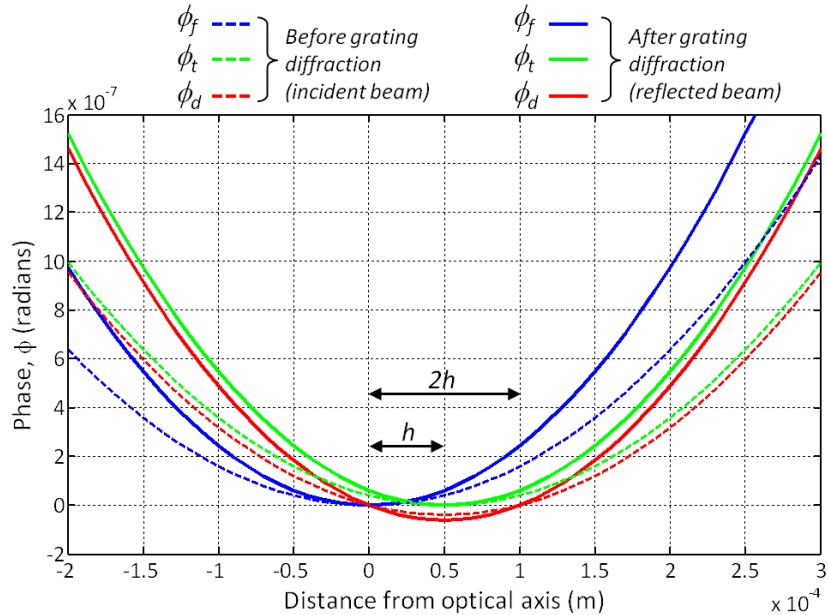}
\caption{Beam phase (wave front)
    for a fundamental non-translated
    beam (solid blue), translated beam (solid green) and a modally
    decomposed beam (solid red). For comparison, the dashed lines
    represent the corresponding beams before grating diffraction. The
    following parameters are assumed: $h=0.05$\,mm, $\lambda =
    10^{-6}$\,m, $z=0.5$\,m, $\omega_{0_i} = 10$\,mm and
    $\omega_{0_r}^x = 8.99$\,mm.}
\label{phase}
\end{figure}

Note that for the incident beam, the waist size $\omega_{0_i}$ is
identical in both $x$ and $y$-directions, and for the diffracted beam,
the waist size in the $y$-direction remains unchanged, implying that
$\omega_{0_r}^y = \omega_{0_i}$. From Eq.~(\ref{eq:grating}), we let
$d=1666$\,nm and $\alpha=10^{\circ}$ to give
$\beta_1=27.7^{\circ}$. If $\omega_{0_i}=10$\,mm, then according to
Eq.~(\ref{eqn14}), $\omega_{0_r}^x=8.99$\,mm. Using these specific
values and Eqs.~(\ref{eqn10}) and (\ref{eqn13}), the phase of each
beam after being diffracted by the grating (and at a distance
$z$ away from the waist position) was plotted against $x$ (the radial
distance from the central optical axis), creating the phase
distributions seen in Fig.~\ref{phase}. Compared to the non-displaced
beam (blue solid line), the phase distribution of a geometrically
translated beam (green solid line) displays the same shape and
profile, and is shifted by an amount $h$ along the $x$-axis, clearly
noticeable in Fig.~\ref{phase}. The phase for a modally decomposed
beam (red solid line) exhibits the same distribution profile, also
shifted by $h$, yet an additional shift along the $y$-axis is evident,
giving rise to `negative' phase. This effect can be clarified by
Eq.~(\ref{eqn13}) in Appendix B: the nature of the equation constricts
the phase distribution for the modally decomposed beam such that the
phase equals zero when $x=0$ and $x=2h$, and hence the reason for the
overlap of the red and blue traces at $x=0$ where there is zero phase.

Subsequently, we find that increasing values of $h$ leads to further
phase deviation of the modally decomposed beam from the geometrically
translated beam. This is simply due to a violation in the
approximation $h/\omega_0 \ll 1$, which was used to obtain
Eq.~(\ref{eqn8}) and in turn Eq.~(\ref{eqn13}).

For comparison, the phase distributions for the non-displaced,
geometrically translated and modally decomposed beams
before grating diffraction are also included (dashed
lines). Notice that although the phase of the
non-diffracted beams present a wider distribution,
deviating away from the solid lines when further away from the optical
axis (as one would expect), the general profile of the dashed traces
is similar to the solid traces and displays no phase difference at the
central optical axis of the beams.

\subsection{Analytical results}
Although the beam suffers from astigmatism as a result of grating
diffraction, this does not change the overall wave front 
for a geometrically translated or modally
decomposed beam, relative to a non-displaced beam. 
Within the Gaussian framework (using Eq.~(\ref{eqn9})), we
find that when a beam is displaced (either through a geometrical
translation or modal decomposition) in increments of increasing $h$,
the phase distribution is always the same. For example, for a
geometrically translated beam, the phase at the central optical axis
is always zero, regardless of the value of $h$. We can therefore
verify that our Gaussian-based framework does not include a change in
phase of $2\pi$ in the order of the grating period (i.e. $\Delta x' =
d$) \cite{freise2007phase}, contrary to the requirements given in
Eq.~(\ref{eq:noise}). The absence of the expected phase change is due
to the fact that the geometrical layout in our framework of beam
displacement is not interchangeable with that of a grating
displacement.

It should be noted that the use of only one higher-order mode
to describe the beam shift introduces small phase differences
and a full expansion into more modes would mitigate this. However, in this discussion 
we are aiming at identifying phase terms in the order of $1$ radian
and thus the simple approximation of using only the first higher-order
mode is sufficient.

We now aim to test the validity of the modal decomposition technique
when describing diffracted displaced beams: we analyse the phase
between various orders of beam modes through experimental means and
justify whether adding the higher-order mode to the fundamental mode
is acceptable.

\section{An experimental demonstration of phase and mode independency}\label{sec:expt}
The modal expansion is using the fact that 
different fields in a linear optical system can be computed separately
and then superimposed to calculate the total field.
In order to further validate this method in the context of
diffraction gratings we use a
table-top experiment to establish that after grating diffraction, the
phase of a zero-order (TEM$_{00}$) beam and a first-order (TEM$_{10}$)
beam are the same, thus signifying that the phase of a diffracted beam
is independent of its mode and thus independent of the intensity 
pattern.

\subsection{Experimental setup}
\begin{figure}[htb]
\centering
\includegraphics[width=4in]{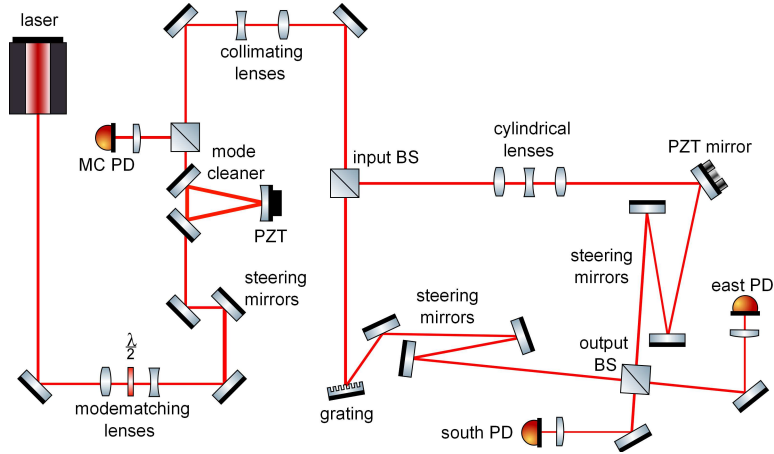}
\caption{\label{MZsetup}Layout of a grating Mach-Zehnder
  interferometer. A square-wave signal is injected into the
  mode-cleaner, allowing the instrument to lock to zero-order and
  first-order mode resonances alternately. One arm of the
  interferometer accommodates the diffraction grating, whilst the
  length of the other arm is subjected to tiny fluctuations to create
  interfering fringe signals at the output.}
\end{figure}

A grating Mach-Zehnder interferometer was developed and used to
distinguish the phase between zero-order and first-order modes, as
depicted in Fig.~\ref{MZsetup}. The laser beam was guided through a
series of modematching lenses and steering mirrors before entering the
triangular mode-cleaner (MC). The MC was tuned by means of a
piezoelectric transducer (PZT), the position of which altered the
round-trip distance of the circulating light inside the MC, and
thereby allowing any chosen mode to resonate and pass through. The
beam was split into two arms of equal lengths via an input
beamsplitter (BS). A ruled reflective diffraction grating was located
in one arm, with a grating period of $d=1666.7$\,nm and aligned to
reflect in the first diffraction order. Note that for the purpose of
this experiment, the grating was fixed in position and was not
translated in any direction. The other arm contained a second PZT,
providing a slight modulation to the arm-length. Both beams recombined
and interfered at the output BS, creating a pair of output beams. Each
output beam consisted of two superimposed beams: grating arm in
transmission plus PZT arm in reflection (denoted as the `east port' in
Fig.~\ref{MZsetup}), and grating arm in reflection plus PZT arm in
transmission (designated as the `south port').

\subsection{Dual-mode locking system}
\begin{figure}
\centering
\includegraphics[width=5in]{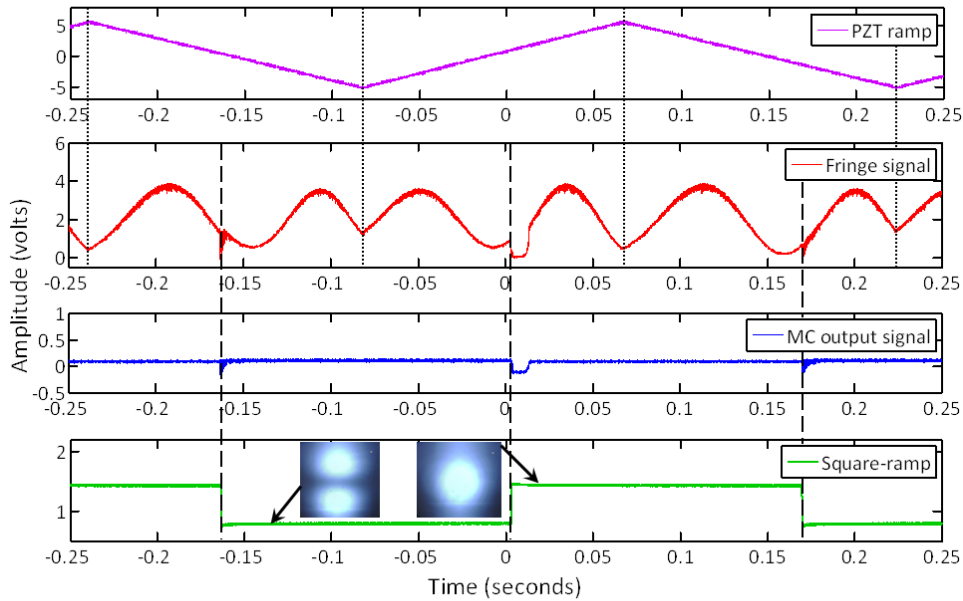}
\caption{\label{fringe}Interference fringe signal during
  mode-switching. From top to bottom: PZT modulation signal in one of
  the Mach-Zehnder arms (purple); fringe signal due to interference at
  the east port (red); output signal from the MC (blue); square-wave
  signal applied to the MC to ramp between modes (green), where the
  maximum and minimum part of the signal correspond to the zero-order
  and first-order mode resonances respectively. As the system switches
  between the modes, the waveform of the fringe signal continues
  undisturbed - this effect is only achieved when the phase of each
  mode is the same. Note that the slight fluctuations visible in the
  fringe signal (and simultaneously in the output of the MC) are due
  to the stabilisation effects of the electronics when locking to each
  mode.}
\end{figure}

The MC operates using feedback control based on the Pound-Drever-Hall
scheme \cite{drever1983laser}. The zero-order and first-order modes
resonating within the MC were isolated and enhanced using a
combination of steering mirrors. The distance between the resonant
peaks of the two modes determined the amplitude of a square-wave
signal (which in this case was 2.1\,V), which in turn was injected
into the MC. The square-wave signal caused the PZT to jump back and
forth between two very specific positions, coinciding with the
zero-order and first-order mode resonances, as revealed by the green
trace in Fig.~\ref{fringe} (set at a frequency of 3\,Hz). As a result, a stable lock to the two
alternating modes was successfully achieved, signified by the emission
of a high and constant level of light power from the MC (blue trace in
Fig.~\ref{fringe}).

In response to the PZT modulation in the Mach-Zehnder arm (purple
trace in Fig.~\ref{fringe}), the superimposed output beams undergo
constructive and destructive interference periodically, and
photodetectors (PDs) situated at each output port detect a fringe
pattern of light and dark bands. Note that when the east port PD
senses constructive interference, the south port PD observes
destructive interference, and vice versa. The red trace in
Fig.~\ref{fringe} is the resulting wave formed from the interference
fringes as detected by the south port PD.

A number of breaks within the fringe waveform are clearly noticeable
in Fig.~\ref{fringe}, corresponding with the vertical dashed lines. We identified that these disruptions coincide
precisely with the moment when the square-wave signal jumps between the
modes. 
Small additional distortions can also be seen at these times, and are caused by the 
temporary disruption of the servo control due to the jump. 
The breaks in fringe symmetry are
due to the change in direction by the PZT mirror in the Mach-Zehnder
arm, hence the correlation with the peaks and troughs of the
triangular wave, indicated by the vertical dotted lines in
Fig.~\ref{fringe}.

\subsection{Experimental results}
If the phase in the zero-order and first-order modes were different,
one would expect a noticeable shift in the phase of the
fringe signal when the modes switched, i.e., at the instant when the
square-wave ramp stepped either up or down. Instead, the results in
Fig.~\ref{fringe} (red trace) show a continuous and unbroken fringe
signal during mode-switching, with no obvious deviation from the
general waveform. The absence of a shift in phase in the fringe signal
confirms the absence of a significant phase difference between both the zero-order and
first-order modes after grating diffraction. This supports the assumption 
that the phase of a diffracted beam is independent of the beam shape.

\section{A rigorous simulation tool for beam diffraction}\label{sec:simulation}
Through the use of simulation techniques, we verify with greater
certainty whether or not the phase of a diffracted beam is independent
of its shape, and we subsequently aim to compare the phase profile of
a diffracted beam when applying the modal decomposition method with
the phase profile of an actual beam displacement. The simulation
relies on a comprehensive technique known as a Finite-Difference
Time-Domain (FDTD) analysis, which can be utilised to create a
powerful two-dimensional simulation tool to solve Maxwell's equations
rigorously in the time-domain
\cite{TafloveHagness200506,yee1966numerical}. For our purposes, an
FDTD implementation was developed \cite{brown2013invariance} to
investigate how beam/grating displacements coupled into the phase of
diffraction orders. A key element is that the FDTD analysis allows for
the simulation of Gaussian beams, which is not the case for other
significant approaches, such as the Rigorous Coupled-Wave Analysis
(RCWA), based on the planewave approximation. The FDTD analysis
enabled us to simulate a diffraction grating and propagate beams with
orders of various mode, and thereby study the phase changes created
when the beam or grating is displaced. The main objective is to
understand why the periodic change in phase of $2\pi$ radians (a
definitive outcome for a displaced beam in a geometric sense) is
absent from the analytical model for a modally decomposed beam in
Section~\ref{sec:theory}.

\subsection{Main parameters and optical layout}
The properties of the modelled optical layout and diffraction grating are shown in
Table~\ref{simvalues}. The cell size, $\Delta$, determines the
resolution of the simulation space - a smaller cell size results in a
higher resolution. For our purpose, the simulation considers only the
$m=0$ and $m=\pm 1$ diffraction orders, and the diffraction angle is
set at $\pm 45^{\circ}$ so as to minimise anisotropic dispersion.

\begin{table}[htb]
\renewcommand{\arraystretch}{1}
\begin{center}
\caption{Parameter values used to simulate Gaussian beam diffraction by a grating.}
\begin{tabular}{ r  l }
\hline
Parameter name & Parameter value\\
\hline
Cell size, $\Delta$ & 25\,nm\\ 
Simulation dimension & 1300$\Delta$ $\times$ 1800$\Delta$ = 32.5\,$\mu$m $\times$ 45\,$\mu$m\\
Wavelength, $\lambda$ & 1064\,nm\\
Grating period, d & 60$\Delta$ = 1500\,nm\\
Beam propagation distance, z & 1200$\Delta$ = 30\,$\mu$m\\
Radius of beam waist, $\omega_0$ & 100$\Delta$ = 2500\,nm\\
\hline
\end{tabular}
\label{simvalues}
\end{center}
\end{table}

The concept of the simulation is as follows: a Gaussian beam is
propagated through a diffraction grating to create $m=0$ and $m=\pm 1$
diffraction orders. For simplicity, a transmission grating is used
instead of a reflection grating - this avoids the need to model a
reflective coating, while the results regarding the phase changes
remain the same.
 
\begin{figure}[htb]
\centering
\begin{minipage}[b]{.48\textwidth}
\centering
\includegraphics[width=6cm]{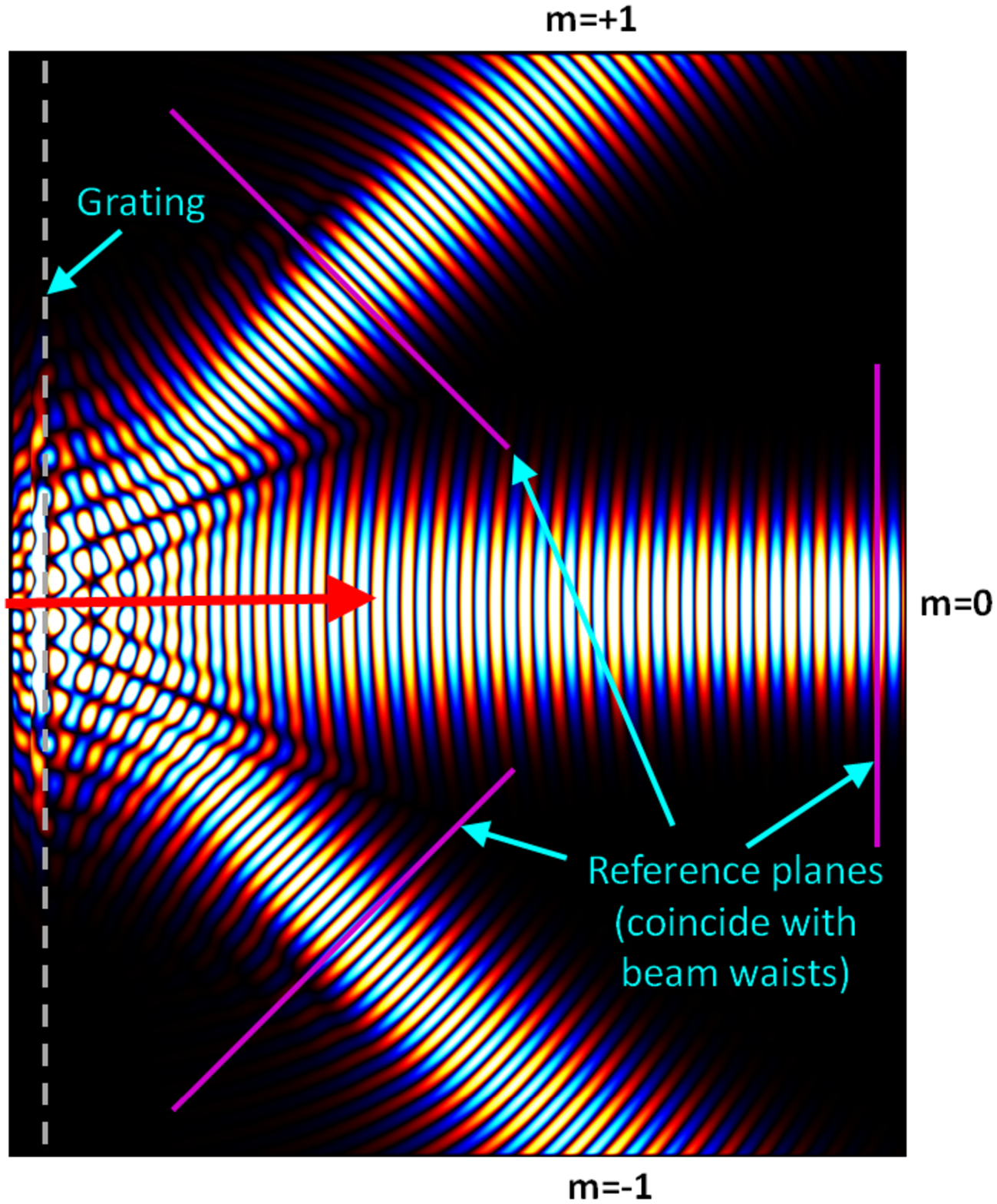}
\end{minipage}
\hspace{0.1cm}
\begin{minipage}[b]{.48\textwidth}
\centering
\includegraphics[width=6cm]{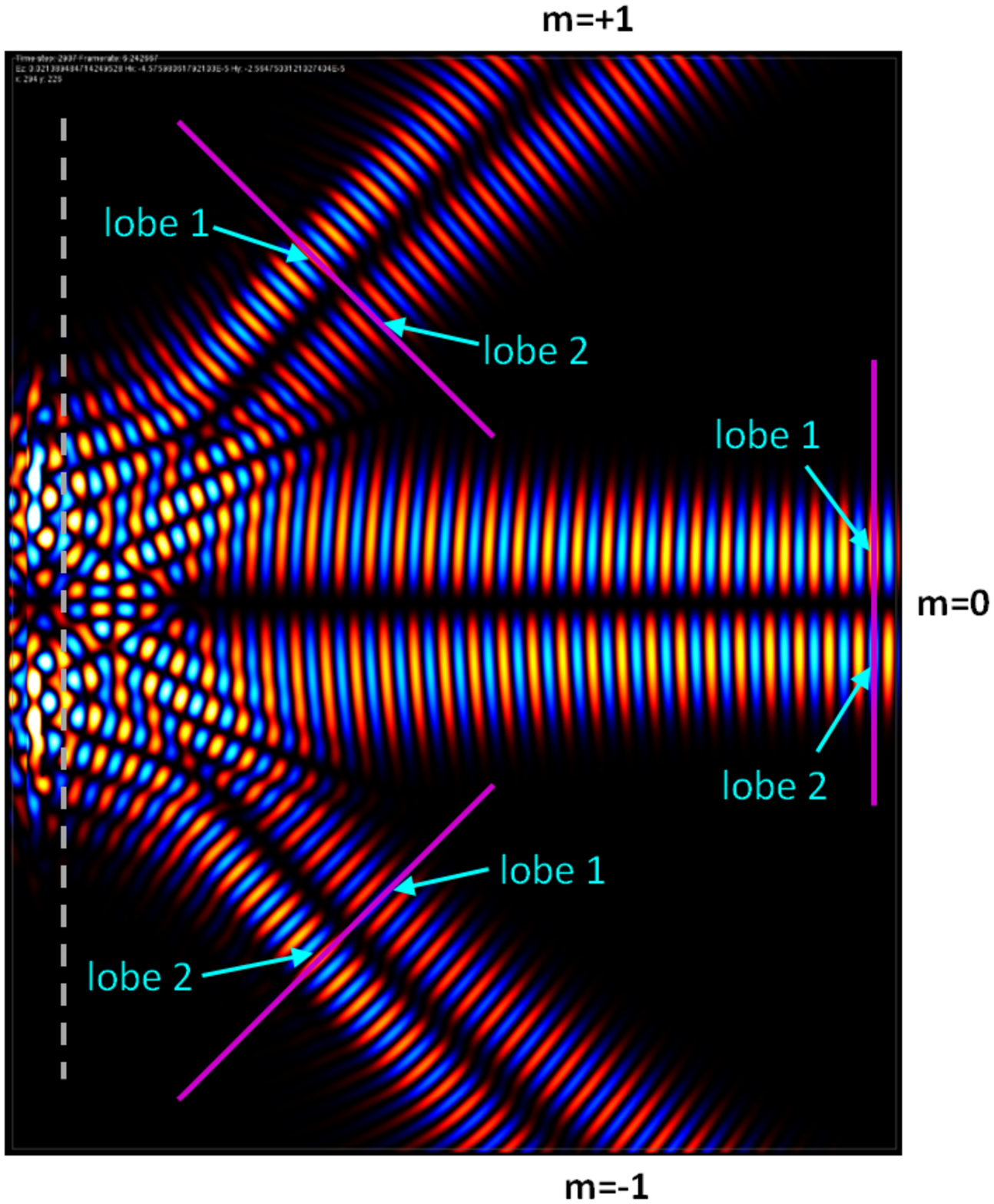}
\end{minipage}
\caption{Diffraction patterns formed by a TEM$_{00}$ beam (left) and a
  TEM$_{10}$ (right). The beams propagate through the diffraction
  grating (dashed grey lines), and the phase for each diffraction
  order, $m=0$ and $m=\pm 1$, is measured at the reference planes
  (solid pink lines).}
\label{prop}
\end{figure}

Figure~\ref{prop} presents an instantaneous image of a
diffraction pattern for a TEM$_{00}$ beam (left), and a TEM$_{10}$
beam (right). The beam travels in the direction indicated by the red
arrow, and the stationary grating is shown by the dashed grey lines.

After diffraction, each beam continues to propagate until they
encounter reference planes, visible as straight, solid pink lines in
Fig.~\ref{prop}. Along each reference plane, a probe sits at the point
of maximum intensity and measures the phase of the passing beam. Since
a TEM$_{10}$ beam consists of two intensity spots (or lobes), each
reference plane contains two probes for phase measurement, as
indicated in Fig.~\ref{prop} (right). We positioned the reference
planes to coincide with the beam waist in all diffraction orders so as
to avoid any extra Gouy phase effects in beams of higher order
modes. For measurements involving grating displacements, the distance
of the grating period $d$ was divided into 40 steps, and the grating
was translated incrementally by a distance of $d/40$ each time. The
total displacement of the incident beam (or grating) was therefore
equivalent to a distance of one grating period ($1.5\,\mu$m), i.e.,
$\Delta x' = d$. At each grating position, the probes recorded the
phase and the simulation was run for some time to allow for better
averaging. The simulation was executed in order to explore two
different scenarios:

\begin{enumerate}
	\item Displacing the incident beam (or grating) vertically using a
    TEM$_{00}$ beam. Note that the context of beam displacement and
    grating displacement are interchangeable if the geometrical layout
    is the same. However, when the beam is displaced in the
    simulation, the reference planes do not follow suit and remain
    static (see Fig.~\ref{prop}). For this reason, it is necessary to
    also translate the reference planes vertically, simultaneously
    with beam displacement, ensuring that the geometry of the layout
    with respect to the grating is consistent.
	\item Modal decomposition, adding a TEM$_{10}$ beam gradually to a
    TEM$_{00}$ beam (thereby reproducing a vertical beam displacement,
    in accordance with Eq.~(\ref{eqn8})). This scenario introduces two
    possibilities: (a) moving the reference planes simultaneously in a
    vertical direction, for reasons of consistency as described in the
    first scenario, and (b) without moving the reference planes. The
    latter raises an interesting issue: given the fact that
    simulations based on modal decomposition are unable to
    automatically change coordinate systems, it follows that the
    reference planes (along which the phase is measured) also remain
    static during beam displacement. We therefore wish to investigate
    both possibilities (a) and (b).
\end{enumerate}

\subsection{Simulation results}
\begin{figure}[htb]
\centering
\includegraphics[width=4.5in]{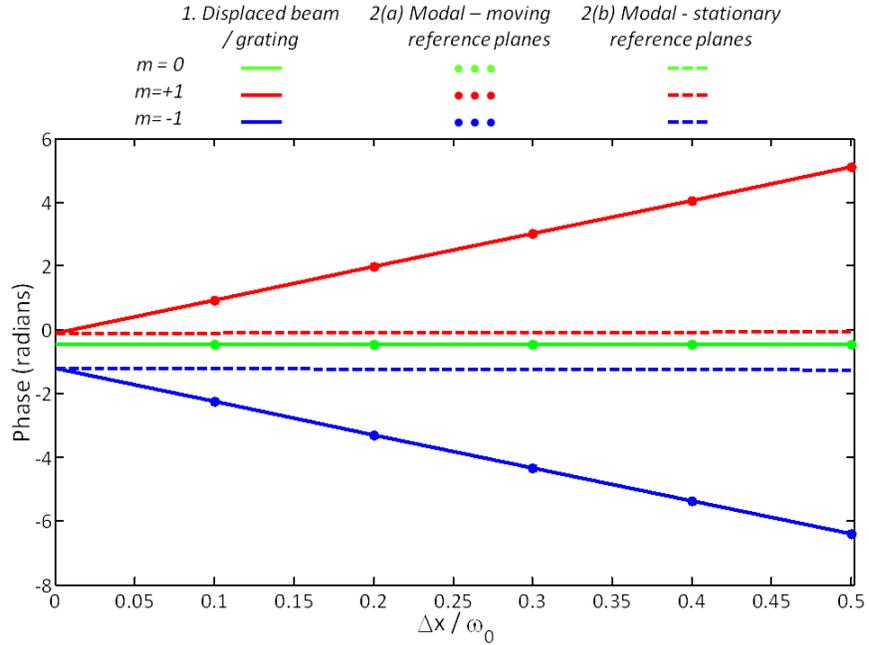}
\caption{Phase changes for a displaced beam after grating diffraction,
  as measured in the $m=0$ and $m=\pm 1$ diffractive orders. Three
  cases are considered: beam/grating displacement (solid), modal
  decomposition with reference planes adjusted vertically (dotted),
  and modal decomposition with fixed reference planes (dashed). Note
  that the green dashed line is coincident with the green solid line.}
\label{phaseplot}
\end{figure}

The change in phase for the three separate cases outlined above in 1,
2(a) and 2(b) are presented in Fig.~\ref{phaseplot}. Each diffraction
order is indicated by the following trace colours: green ($m=0$), red
($m=+1$) and blue ($m=-1$). The various line styles represent each
scenario: beam/grating displacement (solid), modally decomposed with
vertical movement of reference planes (dotted) and modally decomposed
without movement of reference planes (dashed).

\subsubsection{Beam/grating displacement}
During displacement of the grating (or beam) using a TEM$_{00}$ mode
beam, the phase remains constant for $m=0$ (solid green). For $m=+1$
(solid red), the phase gradually increases, and for $m=-1$ (solid
blue), the phase decreases. In each case, the phase undergoes a linear
change of $2\pi$ radians over a total displacement of $d$. The
opposite sign of the slopes are accounted for by the direction of the
grating or beam translation; the optical path length increases in one
of the first diffraction orders and decreases in the other. The
profile of these traces are as predicted by Eq.~(\ref{eq:noise}).

\subsubsection{Modal decomposition - reference planes translated vertically}
When a TEM$_{10}$ mode is added to a TEM$_{00}$ mode beam, the result
is a TEM$_{00}$ beam displaced along the vertical axis (parallel to
the grating). As a higher portion of TEM$_{10}$ is added, the
resulting beam experiences a further displacement. Repositioning the
reference planes along the vertical axis simultaneously with the
resulting beam displacement ensures a consistent geometric layout with
respect to the grating.

During the addition of the higher-order mode, the phase is seen to be
unchanging for $m=0$ (dotted green), and the phase in the $m=+1$
(dotted red) and $m=-1$ (dotted blue) either increase or decrease,
respectively. The phase profiles exhibited in this scenario are highly
comparable with the profiles from the first scenario, and we can
confidently conclude that the addition of a first-order mode to a
zero-order mode beam is an accurate description for small beam or
grating displacements in terms of the phase changes produced.

\subsubsection{Modal decomposition - stationary reference planes}
A TEM$_{10}$ mode beam was added to a TEM$_{00}$ mode beam and
produced a TEM$_{00}$ beam displaced along the vertical axis, as
described before. However, since the reference planes remained static,
after each addition of the higher-order mode (equivalent to each step
of a beam or grating translation), the total optical path length
between the grating and the reference plane increased in one of the
first diffractive orders and decreased in the other (depending on the
direction of the resulting beam displacement). In fact, fixing the
reference planes in one position exactly compensates for the
previously seen linear phase profile, hence we obtain flat lines for
$m=0$ (dashed green, which is not visible because it exactly coincides
with the solid green trace), $m=+1$ (dashed red) and $m=-1$ (dashed
blue).

It is important to note that optical simulation tools used to study
alignment issues, such as \textsc{Finesse}, are restricted in that computations
can only be carried out in one coordinate system. It is for this
reason that these simulation tools rely on the modal decomposition
method for replicating beam displacements. However, the single
coordinate system also means that the reference planes are also forced
to remain stationary, as described in the scenario 2(b). 
We have shown here that in this situation the phase change due to 
beam translation on a diffraction grating is not contained in the
usual model and we recommend a manual injection of this
phase into the analytical model and into existing
Gaussian-based simulation tools in order to obtain an accurate
description of the beam when interacting with diffractive elements.

\section{Conclusion}
We have successfully developed a framework for describing the phase
effects due to beam translations on diffractive elements, incorporating a Gaussian model. We
analysed the modal model for a diffracted Gaussian beam and determined
that the usual modal model does not account for the phase change for
translated or misaligned beams. Using an experimental setup
we demonstrated the absence of phase changes between a diffracted
zero-order beam and first-order beam. Consequently, we confirmed that
the phase of a diffracted Gaussian beam is completely independent of
the beam shape, and that the inclusion of higher-order modes has no
effect on the overall phase. Using a dedicated simulation tool, we examined
the phase changes in the first diffraction order for two scenarios:
firstly during beam/grating displacement for a zero-order beam and
secondly by modal decomposition. We confirmed that the phase changes
resulting from grating displacement were in agreement with a pure
geometric planewave consideration. It is essential that the reference
planes, and therefore the coordinate system, can be moved in order to
obtain the correct phase measurements when reproducing beam
displacements. In a fixed coordinate system required for the common
application of modal models, this can accounted for
most easily by adding 
the extra phase change relating to the change in the geometric path
length when a beam is displaced. This is valid for analytical and
numerical models, and we suggest implementing this procedure to
Gaussian-based simulations relying on modal decomposition, such as
\textsc{Finesse}.

\appendix
\section*{Appendix A: Mathematical derivation of a Gaussian beam}
\setcounter{equation}{0}
\renewcommand{\theequation}{A{\arabic{equation}}}

We present a mathematical description of Gaussian beams in more detail
here. Hermite-Gauss modes have the general form:

\begin{equation}
\label{eqn1}
E(x,y,z) = \sum\limits_{nm} a_{nm}(x,y,z) u_{nm}(x,y,z) e^{-ikz}.
\end{equation}

The normalised Hermite-Gauss function $u_{nm}(x,y,z)$ describes the
transverse spatial distribution of the beam as it varies slowly with
$z$ and is defined as:
\begin{eqnarray}
\label{eqn2}
u_{nm}(x,y,z) &=& \left(2^{n+m-1}n!m!\pi\right)^{\frac{1}{2}} \frac{1}{\omega(z)} \textrm{exp} \big(i(n+m+1)\Psi(z)\big) H_n\left(\frac{\sqrt{2}x}{\omega(z)}\right)\nonumber\\
&\times& H_m\left(\frac{\sqrt{2}y}{\omega(z)}\right) \textrm{exp} \left(-i \frac{k(x^2+y^2)}{2R_C(z)} - \frac{x^2+y^2}{\omega^2(z)}\right),
\end{eqnarray}

where $H_n$ and $H_m$ are Hermite polynomials, $\omega(z)$ is the beam
size, $R_C(z)$ is the radius of curvature of the beam wavefronts. The
Gouy phase is specified as $\Psi(z) = \textrm{arctan} [(z-z_0)/z_R]$,
with $z_R$ being the Rayleigh range. Unless otherwise specified, the
beam waist, $\omega_0$, will always be located at the grating,
i.e. where $z=z_0$.

In general, an offset beam is displaced in both the $x$ and $y$
directions. Due to the symmetry of the system, we consider the
displacement of the beam for only one degree of freedom, along the
$x$-axis. The normalised Hermite-Gauss function, $u_n(x,z)$ in $x$
becomes:
\begin{equation}
\label{eqn3}
u_n(x,z) = \left(\frac{2}{\pi}\right)^{\frac{1}{4}} \left(\frac{\textrm{exp}\big(i(2n+1)\Psi(z)\big)}{2^nn!\omega(z)}\right)^{\frac{1}{2}} H_n\left(\frac{\sqrt{2}x}{\omega(z)}\right) \textrm{exp} \left(-i \frac{kx^2}{2R_C(z)} - \frac{x^2}{\omega^2(z)}\right).
\end{equation}

At the waist, the Gouy phase is zero. In addition, $R_C \rightarrow
\infty$ and therefore the $R_C$ term in Eq.~(\ref{eqn3}) can be
ignored. Since $H_0=1$, a zero-order mode where $n=0$ can be described
at the waist position in the following form:

\begin{equation}
\label{eqn4}
u_0(x,z_0) = \left(\frac{2}{\pi}\right)^{\frac{1}{4}} \frac{1}{\sqrt{\omega_0}} \textrm{exp} \left(- \frac{x^2}{\omega_0^2}\right).
\end{equation}

A first-order mode with $n=1$ at the waist position is given as:

\begin{equation}
\label{eqn5}
u_1(x,z_0) = \left(\frac{2}{\pi}\right)^{\frac{1}{4}} \frac{1}{\sqrt{2\omega_0}} \left(\frac{2\sqrt{2}x}{\omega_0}\right) \textrm{exp} \left(- \frac{x^2}{\omega_0^2}\right).
\end{equation}

Using Eqs.~(\ref{eqn4}) and (\ref{eqn5}), we obtain a simple
relationship between the zero-order and first-order modes at the beam
waist:

\begin{equation}
\label{eqn6}
u_1(x,z_0) = \frac{2x}{\omega_0} u_0(x,z_0).
\end{equation}

\section*{Appendix B: Derivation of phase terms}
\setcounter{equation}{0}
\renewcommand{\theequation}{B{\arabic{equation}}}

In Section~\ref{phaseterms}, the phase, $\theta$, is described for a
stationary fundamental beam (f), a translated beam (t) and a modally
decomposed beam (d). The common factor of $kz - \frac{1}{2}\Psi$ in
Eq.~(\ref{eqn9}) can be omitted to leave $\phi_{f,t,d}$, defined for
each individual beam as follows:

\begin{equation}
\label{eqn10}
\phi_f = \frac{kx^2}{2R_C}, \qquad \phi_t = \frac{k(x-h)^2}{2R_C}, \qquad \phi_d = \frac{kx^2}{2R_C} - \varphi.
\end{equation}

\vspace{1mm}

The $\varphi$ term in Eq.~(\ref{eqn10}) arises from the fact that the
beam in Eq.~(\ref{eqn8}) is a superposition of modes. By expanding and
simplifying Eq.~(\ref{eqn8}), we find that the remaining factor
consists of a sum of terms, which is subsequently treated as a complex
number. We use the relation $e^{i\varphi} = (\textrm{cos} \, \varphi
+i \, \textrm{sin} \, \varphi)$ to reach the expression:

\begin{equation}
\label{eqn13}
\varphi = \textrm{arctan} \left(\frac{\textrm{sin} \, \Psi}{\textrm{cos} \, \Psi + \left(\frac{\omega \, \omega_0}{2xh}\right)}\right).
\end{equation}

\end{document}